\documentclass[12pt]{article}

\usepackage[totalheight = 23cm, totalwidth = 17cm]{geometry}
\usepackage{amssymb,amsmath,amsfonts,amsbsy,graphicx}
\usepackage{bm}

\def\mathbi#1{\textbf{\em #1}}

\newcommand{\mpl}{m_{\rm Pl}}
\newcommand{\fnl}{f_{\rm NL}}

\newcommand{\calO}{{\cal O}}

\newcommand{\calR}{{\cal R}}

\begin{document}

\begin{titlepage}

\rightline{\footnotesize{APCTP-Pre2015-004}} \vspace{-0.2cm}
\rightline{\footnotesize{YITP-15-10}} \vspace{-0.2cm}

\begin{center}

\vskip 1.0 cm

{\LARGE  \bf  A new parameter in \\ attractor single-field inflation}

\vskip 1.0cm

{\large
Jinn-Ouk Gong$^{a,b}$ \hspace{0.2cm} and \hspace{0.2cm} Misao Sasaki$^{c}$
}

\vskip 0.5cm

\small{\it
$^{a}$Asia Pacific Center for Theoretical Physics, Pohang 790-784, Korea
\\
$^{b}$Department of Physics, Postech, Pohang 790-784, Korea
\\
$^{c}$Yukawa Institute for Theoretical Physics, Kyoto University, Kyoto 606-8502, Japan
}

\vskip 1.2cm

\end{center}

\begin{abstract}

We revisit the notion of slow-roll in the context of general single-field 
inflation. As a generalization of slow-roll dynamics, we consider 
an inflaton $\phi$ in an attractor phase where the time derivative 
of $\phi$ is determined by a function of $\phi$, $\dot\phi=\dot\phi(\phi)$. 
In other words, we consider the case when the number of $e$-folds $N$ 
counted backward in time from the end of inflation is solely a 
function of  $\phi$, $N=N(\phi)$. In this case, it is found that 
we need a new independent parameter to properly describe the dynamics 
of the inflaton field in general, in addition to the standard parameters 
conventionally denoted by $\epsilon$, $\eta$, $c_s^2$ and $s$.
Two illustrative examples are presented to discuss the non-slow-roll 
dynamics of the inflaton field consistent with observations.

\end{abstract}

\end{titlepage}

\newpage
\setcounter{page}{1}

\section{Introduction}

The primordial inflation~\cite{inflation} in the very early 
universe before the onset of the standard hot big bang evolution is
now the leading candidate to explain otherwise extremely finely 
tuned initial conditions, such as the horizon and flatness problems. 
Furthermore, inflation can naturally provide a causal mechanism of producing 
primordial curvature perturbations that should have existed on super-horizon 
scales~\cite{Sasaki:2012ss}. 
These primordial curvature perturbations are predicted to have a nearly scale 
invariant power spectrum and are statistically almost perfectly Gaussian. By 
recent observations including the Planck mission, these properties have been
confirmed with very high accuracy~\cite{planck2015,Ade:2015ava,Ade:2015oja}.

While the inflationary picture itself is more and more supported and 
favoured by recent observations, constructing a realistic and concrete 
model of inflation in the context of particle physics remains an open 
conundrum~\cite{Lyth:1998xn}. In this situation we should be open-minded and
consider a wider, more general possibilities for inflation than the simplest
model where a single, canonically normalized inflaton minimally coupled to 
Einstein gravity drives inflation. Such general theories may well predict 
verifiable new observational signatures such as a slight blue tilt for 
tensor perturbations~\cite{bluetensor} and suppression of the curvature
perturbation on large scales~\cite{White:2014aua}. We may have to take 
these possibilities more seriously, as the simplest possibilities including 
the $m^2\phi^2$ model seem to be not favoured by the new Planck
 data~\cite{Ade:2015oja}.

A caution is in order when we study such general possibilities.
We should keep in mind that many notions we have developed in the
canonical models are not directly applicable to them.
For example, the moment of horizon crossing which is crucial for standard 
single field inflation may not be as important as any other instants
 during inflation. This is because, contrary to the canonical model, the 
curvature perturbation may keep evolving on super-horizon scales until 
the end of inflation~\cite{Sasaki:1998ug,Gordon:2000hv} by e.g. 
the existence of other relevant degrees of freedom, which may reflect 
the signatures of high energy physics~\cite{heavy}. 
In this article, we revisit the term ``slow-roll'' in the context of 
k-inflation type general $P(X,\phi)$ theory
where $X \equiv -g^{\mu\nu}\partial_\mu\phi\partial_\nu\phi/2$~\cite{k-inf}.

The article is organized as follows. In Section~\ref{sec:Pderiv} 
we extend the notion of slow-roll single-field inflation and consider 
the general, attractor phase inflation in the context of $P(X,\phi)$ theory.
In describing the dynamics of the inflaton field, we introduce an 
independent new parameter $p$ [see \eqref{p}] which identically vanishes 
in the canonical single-field model. The new parameter is slow-roll 
suppressed if the inflaton is slow-rolling. However, in the general
case of attractor inflation where the inflaton is may not be slow-rolling,
it may become of order unity.
In Section~\ref{sec:examples} we present two examples to 
illustrate the possibility of the non-slow-roll dynamics consistent with
the current observational constraints. We conclude the paper
in Section~\ref{sec:conc}.

\section{General attractor inflation}
\label{sec:Pderiv}

For $P(X,\phi)$ theory, the matter Lagrangian is given by
\begin{equation}
S_m = \int d^4x \sqrt{-g} P(X,\phi) \, .
\end{equation}
This is the most general single scalar field action with their linear 
derivatives, which includes the standard canonical action $P = X - V$ 
and the Dirac-Born-Infeld type action. We assume that the inflaton is
in an attractor phase, i.e., $\dot\phi$ is determined by a function of 
$\phi$, but $\phi$ is not necessarily slowly evolving, as discussed in 
more detail below. Thus, in particular, we do not consider non-attractor 
inflation~\cite{non-attractor} where the dynamics depends 
both on $\phi$ and $\dot\phi$.

With the above Lagrangian, it is known that the spectral index 
of the curvature perturbation is given by~\cite{k-inf}
\begin{equation}\label{index}
n_\calR-1 = -2\epsilon - \eta - s \, ,
\end{equation}
as well as the running of the spectral index~\cite{Lorenz:2008et}
\begin{equation}\label{running}
\alpha_\calR = -2\epsilon\eta - \frac{\dot\eta}{H} - \frac{\dot{s}}{H} \, ,
\end{equation}
where 
\begin{equation}
\begin{split}
 \epsilon & \equiv -\frac{\dot{H}}{H^2}=\frac{XP_X}{\mpl^2H^2} \, ,
 \\
 \eta & \equiv \frac{\dot\epsilon}{H\epsilon}=-\frac{\ddot H}{H^3\epsilon}+2\epsilon \, ,
 \\
 s & \equiv \frac{\dot{c}_s}{Hc_s} \, ,
\end{split}
\label{parameters}
\end{equation}
with the speed of sound $c_s$ given by
\begin{equation}\label{soundspeed}
c_s^{-2} = 1 + \frac{2XP_{XX}}{P_X} \, .
\end{equation}
In deriving \eqref{index}, it is assumed that $H$ and $c_s$ are slowly varying. 
The constrained value of $n_\calR-1 = 0.968 \pm 0.006$~\cite{Ade:2015oja} 
demands $\epsilon$, $\eta$ and $s$ are all small, barring accidental 
cancellation among them. This situation is usually referred to as the 
``slow-roll'' approximation. It is however quite misleading because the 
smallness of these parameters does not necessarily mean the inflaton is
slowly evolving. This becomes more transparent if we consider the equation of 
motion for $\phi$, which reads~\cite{k-inf}
\begin{equation}\label{eom}
\frac{1}{a^3} \frac{d}{dt} \left( a^3P_X\dot\phi \right)
= \frac{d}{dt}(P_X\dot\phi) +3HP_X\dot\phi = P_\phi \,.
\end{equation}
In the canonical case where $P_X=1$, the smallness of $\epsilon$
and $\eta$ would imply the smallness of the $\ddot\phi$ term
in comparison with $3H\dot\phi$ term, which is the usual slow-roll 
approximation. But the second derivative term may not be negligible in 
the general $P(X,\phi)$ theory a priori.

Let us take another point of view by considering the second order
 component of the comoving curvature perturbation $\calR$. 
In the context of the $\delta{N}$ formalism~\cite{Sasaki:1998ug,deltaN}
 where $N = N(\phi)$, for single field case we can find (see for detail 
Appendix~\ref{app:R2nd})
\begin{equation}\label{2ndR}
\delta N=\calR = \calR_l \left[ 1 
+ \frac{1}{2} \left( \epsilon + \delta \right) \calR_l + \cdots \right] \, ,
\end{equation}
where $\calR_l \equiv -H\delta\phi/\dot\phi$ is the linear component of
 $\calR$ with $\delta\phi$ being evaluated on flat slices at horizon crossing,
and
\begin{equation}
\delta \equiv \frac{\ddot\phi}{H\dot\phi} \, .
\label{defdelta}
\end{equation}
The notion of ``slow-roll'', i.e. slow evolution of the inflaton field 
is thus equivalent to requiring $|\delta|\ll1$. Note that \eqref{2ndR} 
is in fact the second order gauge transformation~\cite{Noh:2004bc} and 
is independent of the structure of the matter sector, so should remain 
valid for $P(X,\phi)$ theory. Only when the kinetic sector is canonical 
we can use the relation $\dot{H} = -X/\mpl^2$ 
and find $\eta = 2(\epsilon+\delta)$, so that the smallness of the 
second order component of $\calR$ in \eqref{2ndR} is guaranteed. 
However, in $P(X,\phi)$ theory, $\dot{H}=-XP_X/\mpl^2$ so that 
in general we have
\begin{equation}\label{eta_PX}
\eta = 2(\epsilon+\delta) + p \, ,
\end{equation}
where we have introduced a new parameter $p$ defined by
\begin{equation}\label{p}
p \equiv \frac{\dot{P}_X}{HP_X} \,.
\end{equation}
Thus the coefficient in front of the second order component of $\calR$ is
not necessarily small.

Note that $p$ may be expressed as
\begin{equation}\label{p2}
p = \delta \left( \frac{1}{c_s^2}-1 \right) + \frac{P_{X\phi}}{HP_X}\dot\phi
 = -3-\delta+\frac{P_\phi}{H\dot\phi P_X} \, ,
\end{equation}
where for the second equality we have used the equation of motion \eqref{eom}. Equating these two expressions for $p$, we can eliminate $HP_X$ and can write $p$ as
\begin{equation}
p = \frac{\left( c_s^{-2}-1 \right)\delta + 2(3+\delta)q}{1-2q} 
\quad \text{where} \quad 
q \equiv \frac{XP_{X\phi}}{P_\phi}\,.
\label{pform}
\end{equation}
This is another useful formula. 
Since $p$ is expressed in terms of the cross derivative $P_{X\phi}$, 
we can see the qualitative dependence of $p$ on how close the
theory is to the canonical form where $P_{X\phi}=0$. 
Explicitly, we can express $p$ as
\begin{equation}\label{pasymp}
p \approx \left\{
\begin{array}{ll}
\left( c_s^{-2}-1 \right) \delta 
+ \calO (q) & \text{for } |q|\ll 1 
\\
-3-\delta + \calO(q^{-1}) & \text{for } |q|\gg 1
\end{array}
\right. \, .
\end{equation}
Thus on general ground we expect that when $P(X,\phi)$ is highly non-canonical, 
we may have $|q|\gg1$, and the slow-roll dynamics of the inflaton field is not 
guaranteed. In fact if $|q|\gg1$, combined with \eqref{eta_PX}, 
it is {\em required} that the non-slow-rollness must be as large 
as $\delta \approx 3$ with $\epsilon$ and $\eta$ being kept small.

Before closing this section, let us reconsider the curvature
perturbation expanded to second order \eqref{2ndR} in 
the context of non-Gaussianity. Conventially a local non-Gaussianity
is represented by the non-linear parameter $\fnl$~\cite{Komatsu:2001rj}
which appears in the expansion as
\begin{equation}
\calR = \calR_l + \frac{3}{5}\fnl\calR_l^2 + \cdots \,.
\label{fnldef}
\end{equation}
For the canonical case, using \eqref{eta_PX}, \eqref{2ndR} reads
\begin{equation}\label{2ndR2}
\calR = \calR_l + \frac{\eta}{4}\calR_l^2 + \cdots \, ,
\end{equation}
which implies
\begin{equation}\label{stdconsistency}
\fnl = \frac{5}{12}\eta \,.
\end{equation}
This is in fact a half of the consistency relation for the squeezed limit 
of the bispectrum~\cite{Maldacena:2002vr}. The remaining half 
$5\epsilon/6$ comes from the intrinsic non-Gaussianity of $\calR_l$, 
which we have not taken into account here. See Appendix~\ref{app:int-nG}
 for detail. However, given that for most inflationary 
models $\epsilon\ll1$, \eqref{stdconsistency} contributes more 
importantly to $\fnl$.

Now, following the same step, from \eqref{eta_PX} we obtain
for $P(X,\phi)$ theory,
\begin{equation}
\fnl = \frac{5}{12} \left( \eta - \frac{p}{2} \right) \, .
\end{equation}
One might expect from \eqref{index} that $p$ could be
expressed in terms of $c_s$ or $s$ at an attractor stage 
where $N=N(\phi)$ in a {\em universal} 
manner\footnote{Note that in general $\fnl$ can be comparable to or 
larger than $\calO(1)$ at a non-attractor stage,
breaking the consistency relation, even for canonical 
models~\cite{non-attractor}, along with other possible peculiar 
signatures~\cite{Gong:2014qga}.} .
However, actually it seems there is no universal relation 
between $p$ and $c_s$ or $s$. That is, essentially $p$ is 
an independent parameter of the $P(X,\phi)$ theory.
In the following, let us see this point more clearly 
in two simple examples.

\section{Slow-roll versus non-slow-roll dynamics: Examples}
\label{sec:examples}

If we assume the slow-roll dynamics of the inflaton, i.e. $|\delta|\ll1$, 
from \eqref{eta_PX} we have to {\em additionally} require 
\begin{equation}
|p| \ll 1 \, ,
\end{equation}
given the smallness of $\eta$. However this is an extra assumption not 
constrained by the current observations on $n_\calR$, and in principle can 
be abandoned, \`a la general slow-roll~\cite{gsr} where the hierarchy 
between slow-roll parameters is not assumed. In this case $|\delta| = \calO(1)$ 
can be cancelled by $p \sim -\delta$, keeping small $\eta$ so that there 
is no conflict with observations. Below we present two opposite examples: 
A trivial case where $|\delta|\ll1$ and a non-trivial case where 
$|\delta|=\calO(1)$.

\subsection{Trivial example}

In simple cases, $\eta$ and $\delta$ 
go together, i.e. when one is small, so is the other. As a very simple 
example in this category, consider 
\begin{equation}\label{trivial}
P(X,\phi) = K(X) - V(\phi) \, .
\end{equation}
This gives $P_{X\phi} = 0$, so $p$ is very simple and is related to 
$\delta$ from \eqref{p2} as
\begin{equation}\label{p_trivial}
p = \left( \frac{1}{c_s^2}-1 \right)\delta = \frac{2XK_{XX}}{K_X}\delta \, ,
\end{equation}
where for the second equality we have used the expression
for the speed of sound:
\begin{equation}
c_s^{-2} = 1+\frac{2XK_{XX}}{K_X}\, .
\end{equation}
Thus unless $c_s^2\ll1$, which is highly constrained from bounds on 
$\fnl$ by Planck~\cite{Ade:2015ava,Ade:2015oja}, 
we have $p = \calO(\delta)$. 
Namely $|\eta| \ll 1$ demands $|\delta| \ll 1$, ensuring slow-roll dynamics. 
Notice that if $K\propto X^n$ ($n\neq1$), $c_s$ is constant and $s = 0$, 
hence the spectral index formula \eqref{index} 
as well as the running \eqref{running} are identical to the 
canonical case. But even in this case $\fnl$ is different from the 
canonical case \eqref{stdconsistency} 
because of the non-vanishing new term \eqref{p_trivial},
though this difference is rather irrelevant since we still 
have $\left|\fnl\right|\ll1$.

\subsection{Non-trivial example}

As a non-trivial example where $|\eta| \ll 1$ and $|\delta| \gtrsim 1$,
let us consider
\begin{equation}
P(X,\phi) = F(\phi)K(X)-V(\phi)
\quad \text{with} \quad 
K(X)=\frac{X_0}{1+\gamma}
\left[\left( \frac{X}{X_0}+1 \right)^{1+\gamma}-1\right]\,,
\label{P-nont}
\end{equation}
where $X_0$ is an arbitrary normalization. 
Note that $K\propto X^{\gamma+1}$
for $X\gg X_0$ while $K\propto X$ for $X\ll X_0$. 
Thus the system reduces to the canonical form when $X\ll X_0$
by appropriately redefining the inflaton field.

Taking the derivatives of \eqref{P-nont}, we find
\begin{align}
P_X &= F(\phi)K_X \quad \text{with} \quad K_X
 = \left( \frac{X}{X_0} +1 \right)^\gamma \, ,
\\
P_{XX} & = F(\phi)K_{XX}\quad \text{with} \quad K_{XX}
=\frac{\gamma}{X_0} \left( \frac{X}{X_0}+1 \right)^{\gamma-1} \, ,
\end{align}
and from \eqref{soundspeed},
\begin{equation}
c_s^{-2} = 1+\frac{2XK_{XX}}{K_X}=1 + 2\gamma 
\frac{\left(X/X_0+1\right)^{\gamma-1}X/X_0}{\left(X/X_0+1\right)^\gamma} \, .
\end{equation}
Note that for $X \gg X_0$, we have a simple result,
\begin{equation}\label{cs2}
c_s^{-2} \approx 1+2\gamma \,.
\end{equation}
Note also that with $\gamma$ being a constant, $s \approx 0$ in this limit.

In the following, let us concentrate on this regime. 
To make the analysis simpler, we assume the time dependence of 
$\phi$ as
\begin{equation}\label{phi_ansatz}
\phi \sim e^{\alpha N} \, ,
\end{equation}
where we are interested in the case when $\alpha$ is not small,
$\alpha\gtrsim\calO(1)$. The consistency of this assumption will
be discussed later. Accordingly we find
\begin{equation}
\begin{split}
\dot\phi & = \alpha H\phi \, ,
\\
X & = \frac{\alpha^2H^2}{2}\phi^2 \, ,
\\
\frac{dX}{dN} & = 2(\alpha-\epsilon) X \, .
\end{split}
\end{equation}

What we want to see is whether $|\delta| \gtrsim 1$ while $|\eta| \ll 1$ 
is possible. For this purpose, let us express $\eta$ in the form,
\begin{equation}\label{eta:model}
\eta = 2\epsilon
 + \frac{1}{F}\frac{dF}{dN} + \frac{1}{XK_X} \frac{d(XK_X)}{dN} \,,
\end{equation}
where we have used \eqref{eta_PX} and \eqref{p}.
Note that $\delta$ is expressed as
\begin{equation}
\delta = \frac{1}{2} \frac{\dot{X}}{HX} = \frac{1}{2X}\frac{dX}{dN}
 = \alpha-\epsilon \,.
\end{equation}
We see that given $\epsilon \ll 1$ the last two terms 
in \eqref{eta:model} should nearly 
cancel each other to ensure small $\eta$.

With \eqref{phi_ansatz} and $X \gg X_0$, we find
\begin{equation}
\frac{1}{XK_X} \frac{d(XK_X)}{dN} \approx 2 (\alpha-\epsilon) (1+\gamma) \, .
\end{equation}
Now let us set
\begin{equation}
\frac{1}{F}\frac{dF}{dN} = -2 (\alpha-\epsilon) (1+\gamma) + \xi \,.
\end{equation}
For the last two terms in \eqref{eta:model} to nearly cancel each other,
we must have $|\xi|\ll1$. This implies
\begin{equation}
F \approx F_0 \left(\frac{\phi}{\phi_0}\right)^{-2(1+\gamma)} \,,
\end{equation}
where we have ignored the corrections of $\calO(\epsilon)$.

In the limit $X\gg X_0$, $\epsilon$ is given by
\begin{equation}
\epsilon \approx \frac{FX^{1+\gamma}}{\mpl^2H^2X_0^\gamma}
 \approx \frac{F_0\alpha^{2(1+\gamma)}}{2^{1+\gamma}} 
\frac{\phi_0^{2(\gamma+1)}H^{2\gamma}}{\mpl^2X_0^\gamma}\, .
\end{equation}
Thus by appropriately choosing the normalization constants, we can readily 
make $\epsilon\ll1$.
 Turning to the equation of motion for $\phi$, \eqref{eom}, we obtain
\begin{equation}
\frac{F_0}{X_0^\gamma} \frac{(\alpha H)^{2(1+\gamma)}}{2^\gamma} 
\left[ 3+\frac{1+2\gamma}{2(1+\gamma)}\xi \right] \frac{dN}{d\phi}
 = -V_\phi \, .
\end{equation}
Then, ignoring $\xi$ as well as the time variation of $H$, 
we can recover the advocated behavior \eqref{phi_ansatz}
with a logarithmic potential,
\begin{equation}
V(\phi) = V_0 + V_1\log\left(\frac{\phi}{\phi_0}\right) \, ,
\end{equation}
upon appropriately choosing $V_1$.

Notice that even for $X \gg X_0$ where we can make simplifications, 
$p$ is not related to the speed of sound $c_s$ \eqref{cs2}. Instead
we find
\begin{equation}
p =\frac{\dot F}{HF}+\frac{\dot K_X}{HK_X}
\approx-2\alpha(1+\gamma)+2\alpha\gamma\approx -2\delta \, ,
\end{equation}
where for the last equality we have used $\alpha \approx \delta$.
We see that $\gamma \approx \left(c_s^{-2}-1\right)/2$ disappears 
from the final result, implying that $p$ is related to 
neither $c_s$ nor $s$. From \eqref{eta_PX} we find
\begin{equation}
\eta = 2(\epsilon+\delta)+p 
\approx 2(\epsilon+\delta)-2\delta \approx 2\epsilon \ll 1 \, ,
\end{equation}
as required, so that there is no conflict with the observational
constraints on $n_\calR$. Further, in this regime the running is 
$\alpha_\calR \approx -8\epsilon^2$ which can be well accommodated
within the current observational bounds. On the other hand we find
\begin{equation}
\fnl = \frac{5}{12} \left( \eta - \frac{p}{2} \right) 
\approx \frac{5}{12} ( \eta + \delta ) 
\approx \frac{5}{12}\alpha \gtrsim \calO(1) \, .
\end{equation}
Note that we have
\begin{equation}
\frac{XP_{X\phi}}{P_\phi} \approx -\frac{\alpha(1+\gamma)}{3-\alpha} \, ,
\end{equation}
so that upon choosing $\alpha \approx \delta \approx 3$ 
the dynamics of the inflaton becomes maximally non-slow-roll,
corresponding the general case of $|q|\gg1$ given in \eqref{pasymp}.

\section{Conclusion}
\label{sec:conc}

We have reconsidered the notion of slow-roll in the context of 
general $P(X,\phi)$ theory in terms of the parameters by which observable
quantities are described. While in the standard single-field inflation
the attractor phase corresponds to the slow-roll regime, they are not 
equivalent in general. Accordingly we have found that we need a new, 
independent parameter $p$ defined by \eqref{p} to properly describe 
the dynamics of the inflaton field.
We have presented two illustrative examples in order to clarify the role 
of the new parameter in the non-slow-roll dynamics of the inflaton field.
In one of the examples, we have shown that we may indeed have a
highly non-slow-roll stage of inflation without violating the current
observational constraints. In other words, in near-future observations 
where the precision and accuracy will become much better, this new parameter 
can be used to perform a new observational test to constrain viable models
of inflation.

\subsection*{Acknowledgements}

JG is grateful to the Yukawa Institute for Theoretical Physics for
 hospitality while this work was under progress.
JG acknowledges the Max-Planck-Gesellschaft, the Korea Ministry of 
Education, Science and Technology, Gyeongsangbuk-Do and Pohang City
 for the support of the Independent Junior Research Group at the 
Asia Pacific Center for Theoretical Physics. 
This work is also supported in part by a Starting Grant through the Basic 
Science Research Program of the National Research Foundation of Korea 
(2013R1A1A1006701) and
the JSPS Grant-in-Aid for Scientific Research (A) No. 21244033.

\newpage

\appendix

\section{Non-linear $\bm\calR$ on attractor}
\label{app:R2nd}

On totally general ground, once the trajectory is in an attractor phase
we can write
\begin{equation}\label{deltaN}
\calR = \delta{N} = \frac{\partial{N}}{\partial\phi}\delta\phi 
+ \frac{1}{2} \frac{\partial^2N}{\partial\phi^2}\delta\phi^2 + \cdots \, ,
\end{equation}
where $\calR$ is the comoving curvature perturbation evaluated at some 
final time, and $\delta\phi$ is the field fluctuation on the initial 
flat slice. We can find the expansion coefficients explicitly as follows.

First, we note $dN=-Hdt$ where the minus sign is due to the
fact that $N(\phi)$ is defined as the number of $e$-folds counted 
{\em backward in time} from the end of inflation to an initial time 
when the value of the inflation field was $\phi$.
Therefore we have
\begin{equation}
\frac{\partial{N}}{\partial\phi} 
= \frac{d{t}}{d\phi}\frac{\partial{N}}{\partial{t}}
= -\frac{H}{\dot\phi} \, .
\end{equation}
Once we have the first coefficient, it is straightforward to 
compute the other ones. The second order coefficient is
\begin{equation}
\frac{\partial}{\partial\phi} \left( \frac{\partial{N}}{\partial\phi} \right) 
= \frac{1}{\dot\phi} \frac{d}{dt}\left( \frac{\partial{N}}{\partial\phi} \right)
= -\frac{\dot{H}}{\dot\phi^2} + \frac{H\ddot\phi}{\dot\phi^3} \, .
\end{equation}

Plugging these coefficients into (\ref{deltaN}), we find
\begin{align}\label{deltaNattractor}
\calR & = -\frac{H}{\dot\phi}\delta\phi 
+ \frac{1}{2} \left( -\frac{\dot{H}}{\dot\phi^2} 
+ \frac{H\ddot\phi}{\dot\phi^3} \right) \delta\phi^2 + \cdots
\nonumber\\
 & = -\frac{H}{\dot\phi}\delta\phi 
+ \frac{1}{2} \left( -\frac{\dot{H}}{H^2}+ \frac{\ddot\phi}{H\dot\phi} \right)
\left( -\frac{H}{\dot\phi}\delta\phi \right)^2 + \cdots \, .
\end{align}
Note that we have {\em not} assumed any particular form for the matter sector.
Thus, using the definitions of $\epsilon$ in \eqref{parameters}
and $\delta$ in \eqref{defdelta} in the main text,
and identifying $\calR_l \equiv -H\delta\phi/\dot\phi$, 
\eqref{deltaNattractor} becomes
\begin{equation}
\calR
 = \calR_l \left[ 1 + \frac{1}{2} (\epsilon+\delta)\calR_l + \cdots \right] \,.
\end{equation}
This is \eqref{2ndR} used in the main text.

\section{Intrinsic non-Gaussianity of $\bm\calR$}
\label{app:int-nG}

We consider the cubic order action of $\delta\phi$ on flat 
slices~\cite{Maldacena:2002vr},
\begin{equation}\label{S3deltaphi}
S_3 = \int d^4x a^3 \left[ -\frac{\dot\phi}{4\mpl^2H}\delta\phi\dot{\delta\phi}^2
 - \frac{\dot\phi}{4\mpl^2H}\delta\phi \frac{\left(\Delta[\delta\phi]\right)^2}{a^2}
 - \dot{\delta\phi}\frac{\delta\phi^{,i}\chi_{,i}}{a^2} + \cdots \right] \, ,
\end{equation}
where we have only presented the leading order terms in slow-roll, 
and $\chi$ is the scalar component of the shift vector given by
\begin{equation}
\frac{1}{a^2}\Delta[\chi]
= \epsilon \frac{d}{dt} \left( -\frac{H}{\dot\phi}\delta\phi \right) 
= -\frac{\dot\phi}{2\mpl^2H}\dot{\delta\phi} 
+ \text{higher order in slow-roll} \, .
\end{equation}
Using the Bunch-Davies mode function in the de Sitter approximation 
with the conformal time $\tau=-1/(aH)$,
\begin{equation}
\delta\phi_k(\tau) = -\frac{\dot\phi}{H}\calR_k(\tau) 
= -\frac{\dot\phi}{H} \frac{iH}{\sqrt{4\epsilon k^3}\mpl} 
\left( 1+ik\tau \right) e^{-ik\tau} \, ,
\end{equation}
we find the three-point correlation function of $\delta\phi$ 
from \eqref{S3deltaphi} as
\begin{align}\label{deltaphi3point3}
\left\langle \delta\phi_{\mathbi{k}_1}\delta\phi_{\mathbi{k}_2}
\delta\phi_{\mathbi{k}_3} \right\rangle 
& \equiv (2\pi)^3 \delta^{(3)}(\mathbi{k}_1+\mathbi{k}_2+\mathbi{k}_3) 
B_{\delta\phi}(k_1,k_2,k_3) 
\nonumber\\
& = (2\pi)^3 \delta^{(3)}(\mathbi{k}_1+\mathbi{k}_2+\mathbi{k}_3) 
\frac{-H^4}{\sqrt{2\epsilon}\mpl} \frac{1}{(k_1k_2k_3)^3} \frac{\epsilon}{4} 
\nonumber\\
& \qquad \times \left[ -\frac{k_1^3+k_2^3+k_3^3}{2} 
+ \frac{k_1\left( k_2^2+k_3^2 \right) + \text{2 perm}}{2} 
+ \frac{4\left( k_1^2k_2^2 + \text{2 perm} \right)}{k_1+k_2+k_3} \right] \, .
\end{align}

Since $\delta\phi=-(\dot\phi/H)\calR$ at leading order, the intrinsic 
bispectrum for the comoving curvature perturbation is given by
\begin{align}
B_\calR(k_1,k_2,k_3) & = -\frac{H^3}{\dot\phi^3} B_{\delta\phi}(k_1,k_2,k_3)
\nonumber\\
& = \frac{H^4}{16\epsilon\mpl^4} \frac{1}{(k_1k_2k_3)^3} 
\left[ -\frac{k_1^3+k_2^3+k_3^3}{2} + \frac{k_1\left( k_2^2+k_3^2 \right)
 + \text{2 perm}}{2} 
+ \frac{4\left( k_1^2k_2^2 + \text{2 perm} \right)}{k_1+k_2+k_3} \right] \, .
\end{align}
Taking the squeezed limit, say, $k_3\to0$, one can read off
the non-linear parameter $\fnl$ and find
\begin{equation}
\fnl = \frac{5}{6}\epsilon \, .
\end{equation}
Thus, we see that using the $\delta{N}$ formalism alone, we cannot fully 
find the consistency relation. From the beginning the $\delta{N}$ 
formalism captures only the super-horizon evolution, which gives
a half of the consistency relation $\fnl=5\eta/12$. The remaining half, 
$\fnl=5\epsilon/6$, is due to the intrinsic non-Gaussianity, which we can 
find from the cubic order action.


\begin{thebibliography}{99}


\bibitem{inflation}
%\cite{Sato:1980yn}
%\bibitem{Sato:1980yn} 
  K.~Sato,
  %``First Order Phase Transition of a Vacuum and Expansion of the Universe,''
  Mon.\ Not.\ Roy.\ Astron.\ Soc.\  {\bf 195}, 467 (1981)~;
  %%CITATION = MNRAA,195,467;%%
%\cite{Guth:1980zm}
%\bibitem{Guth:1980zm}
  A.~H.~Guth,
  %``The Inflationary Universe: A Possible Solution To The Horizon And Flatness
  %Problems,''
  Phys.\ Rev.\  D {\bf 23}, 347 (1981)~;
  %%CITATION = PHRVA,D23,347;%%
%\cite{Linde:1981mu}
%\bibitem{Linde:1981mu}
  A.~D.~Linde,
  %``A New Inflationary Universe Scenario: A Possible Solution Of The Horizon,
  %Flatness, Homogeneity, Isotropy And Primordial Monopole Problems,''
  Phys.\ Lett.\  B {\bf 108}, 389 (1982)~;
  %%CITATION = PHLTA,B108,389;%%
%\cite{Albrecht:1982wi}
%\bibitem{Albrecht:1982wi}
  A.~Albrecht and P.~J.~Steinhardt,
  %``Cosmology For Grand Unified Theories With Radiatively Induced Symmetry
  %Breaking,''
  Phys.\ Rev.\ Lett.\  {\bf 48}, 1220 (1982).
  %%CITATION = PRLTA,48,1220;%%


%\cite{Sasaki:2012ss}
\bibitem{Sasaki:2012ss} 
see e.g.,
 M.~Sasaki,
 ``Inflation and Birth of Cosmological Perturbations,''
  arXiv:1210.7880 [astro-ph.CO].
  %%CITATION = ARXIV:1210.7880;%%
in {\it General Relativity, Cosmology and Astrophysics
Perspectives 100 years after Einstein's stay in Prague},
Fundamental Theories of Physics, Vol. 177,
J. Bicak, T. Ledvinka (Eds.), Springer, Switzerland 2014, 
DOI: 10.1007/978-3-319-06349-2.


\bibitem{planck2015}
%\cite{Adam:2015rua}
%\bibitem{Adam:2015rua} 
  R.~Adam {\it et al.}  [Planck Collaboration],
  %``Planck 2015 results. I. Overview of products and scientific results,''
  arXiv:1502.01582 [astro-ph.CO]~;
  %%CITATION = ARXIV:1502.01582;%%
%\cite{Planck:2015xua}
%\bibitem{Planck:2015xua} 
  P.~A.~R.~Ade {\it et al.}  [Planck Collaboration],
  %``Planck 2015 results. XIII. Cosmological parameters,''
  arXiv:1502.01589 [astro-ph.CO].
  %%CITATION = ARXIV:1502.01589;%%


%\cite{Ade:2015ava}
\bibitem{Ade:2015ava} 
  P.~A.~R.~Ade {\it et al.}  [Planck Collaboration],
  %``Planck 2015 results. XVII. Constraints on primordial non-Gaussianity,''
  arXiv:1502.01592 [astro-ph.CO].
  %%CITATION = ARXIV:1502.01592;%%
  
  
%\cite{Ade:2015oja}
\bibitem{Ade:2015oja} 
  P.~A.~R.~Ade {\it et al.}  [Planck Collaboration],
  %``Planck 2015. XX. Constraints on inflation,''
  arXiv:1502.02114 [astro-ph.CO].
  %%CITATION = ARXIV:1502.02114;%%


%\cite{Lyth:1998xn}
\bibitem{Lyth:1998xn} 
See e.g. 
  D.~H.~Lyth and A.~Riotto,
%``Particle physics models of inflation and the cosmological density perturbation,''
  Phys.\ Rept.\  {\bf 314}, 1 (1999)
  [hep-ph/9807278].
  %%CITATION = HEP-PH/9807278;%%  


\bibitem{bluetensor}
%\cite{Kobayashi:2010cm}
%\bibitem{Kobayashi:2010cm} 
  T.~Kobayashi, M.~Yamaguchi and J.~Yokoyama,
  %``G-inflation: Inflation driven by the Galileon field,''
  Phys.\ Rev.\ Lett.\  {\bf 105}, 231302 (2010)
  [arXiv:1008.0603 [hep-th]]~;
  %%CITATION = ARXIV:1008.0603;%%
%\cite{Cannone:2014uqa}
%\bibitem{Cannone:2014uqa} 
  D.~Cannone, G.~Tasinato and D.~Wands,
  %``Generalised tensor fluctuations and inflation,''
  JCAP {\bf 1501}, no. 01, 029 (2015)
  [arXiv:1409.6568 [astro-ph.CO]]~;
  %%CITATION = ARXIV:1409.6568;%%
%\cite{Cai:2014uka}
%\bibitem{Cai:2014uka} 
  Y.~F.~Cai, J.~O.~Gong, S.~Pi, E.~N.~Saridakis and S.~Y.~Wu,
  %``On the possibility of blue tensor spectrum within single field inflation,''
  arXiv:1412.7241 [hep-th].
  %%CITATION = ARXIV:1412.7241;%%


%\cite{White:2014aua}
\bibitem{White:2014aua} 
  J.~White, Y.~l.~Zhang and M.~Sasaki,
  %``Scalar suppression on large scales in open inflation,''
  Phys.\ Rev.\ D {\bf 90}, no. 8, 083517 (2014)
  [arXiv:1407.5816 [astro-ph.CO]].
  %%CITATION = ARXIV:1407.5816;%%


%\cite{Sasaki:1998ug}
\bibitem{Sasaki:1998ug} 
  M.~Sasaki and T.~Tanaka,
  %``Superhorizon scale dynamics of multiscalar inflation,''
  Prog.\ Theor.\ Phys.\  {\bf 99}, 763 (1998)
  [gr-qc/9801017].
  %%CITATION = GR-QC/9801017;%%


%\cite{Gordon:2000hv}
\bibitem{Gordon:2000hv} 
  C.~Gordon, D.~Wands, B.~A.~Bassett and R.~Maartens,
  %``Adiabatic and entropy perturbations from inflation,''
  Phys.\ Rev.\ D {\bf 63}, 023506 (2001)
  [astro-ph/0009131].
  %%CITATION = ASTRO-PH/0009131;%%
  

\bibitem{heavy}
%\cite{Tolley:2009fg}
%\bibitem{Tolley:2009fg} 
  A.~J.~Tolley and M.~Wyman,
  %``The Gelaton Scenario: Equilateral non-Gaussianity from multi-field dynamics,''
  Phys.\ Rev.\ D {\bf 81}, 043502 (2010)
  [arXiv:0910.1853 [hep-th]]~;
  %%CITATION = ARXIV:0910.1853;%%
%\cite{Achucarro:2010jv}
%\bibitem{Achucarro:2010jv} 
  A.~Achucarro, J.~-O.~Gong, S.~Hardeman, G.~A.~Palma and S.~P.~Patil,
  %``Mass hierarchies and non-decoupling in multi-scalar field dynamics,''
  Phys.\ Rev.\ D {\bf 84}, 043502 (2011)
  [arXiv:1005.3848 [hep-th]]~;
  %%CITATION = ARXIV:1005.3848;%%  
%\cite{Achucarro:2010da}
%\bibitem{Achucarro:2010da} 
  A.~Achucarro, J.~-O.~Gong, S.~Hardeman, G.~A.~Palma and S.~P.~Patil,
  %``Features of heavy physics in the CMB power spectrum,''
  JCAP {\bf 1101}, 030 (2011)
  [arXiv:1010.3693 [hep-ph]]~;
  %%CITATION = ARXIV:1010.3693;%%
%\cite{Achucarro:2012sm}
%\bibitem{Achucarro:2012sm} 
  A.~Achucarro, J.~-O.~Gong, S.~Hardeman, G.~A.~Palma and S.~P.~Patil,
  %``Effective theories of single field inflation when heavy fields matter,''
  JHEP {\bf 1205}, 066 (2012)
  [arXiv:1201.6342 [hep-th]].
  %%CITATION = ARXIV:1201.6342;%%  


\bibitem{k-inf}
%\cite{ArmendarizPicon:1999rj}
%\bibitem{ArmendarizPicon:1999rj} 
  C.~Armendariz-Picon, T.~Damour and V.~F.~Mukhanov,
  %``k - inflation,''
  Phys.\ Lett.\ B {\bf 458}, 209 (1999)
  [hep-th/9904075]~;
  %%CITATION = HEP-TH/9904075;%%
%\cite{Garriga:1999vw}
%\bibitem{Garriga:1999vw} 
  J.~Garriga and V.~F.~Mukhanov,
  %``Perturbations in k-inflation,''
  Phys.\ Lett.\ B {\bf 458}, 219 (1999)
  [hep-th/9904176].
  %%CITATION = HEP-TH/9904176;%%


%\cite{Lorenz:2008et}
\bibitem{Lorenz:2008et} 
  L.~Lorenz, J.~Martin and C.~Ringeval,
  %``K-inflationary Power Spectra in the Uniform Approximation,''
  Phys.\ Rev.\ D {\bf 78}, 083513 (2008)
  [arXiv:0807.3037 [astro-ph]].
  %%CITATION = ARXIV:0807.3037;%%


\bibitem{non-attractor}
%\cite{Namjoo:2012aa}
%\bibitem{Namjoo:2012aa} 
  M.~H.~Namjoo, H.~Firouzjahi and M.~Sasaki,
  %``Violation of non-Gaussianity consistency relation in a single field inflationary model,''
  Europhys.\ Lett.\  {\bf 101}, 39001 (2013)
  [arXiv:1210.3692 [astro-ph.CO]]~;
  %%CITATION = ARXIV:1210.3692;%%
%\cite{Chen:2013aj}
%\bibitem{Chen:2013aj} 
  X.~Chen, H.~Firouzjahi, M.~H.~Namjoo and M.~Sasaki,
  %``A Single Field Inflation Model with Large Local Non-Gaussianity,''
  Europhys.\ Lett.\  {\bf 102}, 59001 (2013)
  [arXiv:1301.5699 [hep-th]]~;
  %%CITATION = ARXIV:1301.5699;%%
%\cite{Chen:2013kta}
%\bibitem{Chen:2013kta} 
  X.~Chen, H.~Firouzjahi, M.~H.~Namjoo and M.~Sasaki,
  %``Fluid Inflation,''
  JCAP {\bf 1309}, 012 (2013)
  [arXiv:1306.2901 [hep-th]]~;
  %%CITATION = ARXIV:1306.2901;%%
%\cite{Chen:2013eea}
%\bibitem{Chen:2013eea} 
  X.~Chen, H.~Firouzjahi, E.~Komatsu, M.~H.~Namjoo and M.~Sasaki,
  %``In-in and $\delta N$ calculations of the bispectrum from non-attractor single-field inflation,''
  JCAP {\bf 1312}, 039 (2013)
  [arXiv:1308.5341 [astro-ph.CO]]~;
  %%CITATION = ARXIV:1308.5341;%%
%\cite{Mooij:2015yka}
%\bibitem{Mooij:2015yka} 
  S.~Mooij, G.~A.~Palma and A.~E.~Romano,
  %``Consistently violating the non-Gaussian consistency relation,''
  arXiv:1502.03458 [astro-ph.CO].


\bibitem{deltaN}
%\cite{Starobinsky:1986fxa}
%\bibitem{Starobinsky:1986fxa} 
  A.~A.~Starobinsky,
  %``Multicomponent de Sitter (Inflationary) Stages and the Generation of Perturbations,''
  JETP Lett.\  {\bf 42}, 152 (1985)
  [Pisma Zh.\ Eksp.\ Teor.\ Fiz.\  {\bf 42}, 124 (1985)]~;
  %%CITATION = JTPLA,42,152;%%
%\cite{Sasaki:1995aw}
%\bibitem{Sasaki:1995aw} 
  M.~Sasaki and E.~D.~Stewart,
  %``A General analytic formula for the spectral index of the density perturbations produced during inflation,''
  Prog.\ Theor.\ Phys.\  {\bf 95}, 71 (1996)
  [astro-ph/9507001]~;
  %%CITATION = ASTRO-PH/9507001;%%  
%\bibitem{Gong:2002cx} 
  J.~-O.~Gong and E.~D.~Stewart,
  %``The Power spectrum for a multicomponent inflaton to second order corrections in the slow roll expansion,''
  Phys.\ Lett.\ B {\bf 538}, 213 (2002)
  [astro-ph/0202098].
  %%CITATION = ASTRO-PH/0202098;%%


%\cite{Noh:2004bc}
\bibitem{Noh:2004bc} 
  H.~Noh and J.~c.~Hwang,
  %``Second-order perturbations of the Friedmann world model,''
  Phys.\ Rev.\ D {\bf 69}, 104011 (2004)
  [astro-ph/0305123].
  %%CITATION = PHRVA,D69,104011;%%


%\cite{Komatsu:2001rj}
\bibitem{Komatsu:2001rj} 
  E.~Komatsu and D.~N.~Spergel,
  %``Acoustic signatures in the primary microwave background bispectrum,''
  Phys.\ Rev.\ D {\bf 63}, 063002 (2001)
  [astro-ph/0005036].
  %%CITATION = ASTRO-PH/0005036;%%


%\cite{Maldacena:2002vr}
\bibitem{Maldacena:2002vr} 
  J.~M.~Maldacena,
  %``Non-Gaussian features of primordial fluctuations in single field inflationary models,''
  JHEP {\bf 0305}, 013 (2003)
  [astro-ph/0210603].
  %%CITATION = ASTRO-PH/0210603;%%


%\cite{Gong:2014qga}
\bibitem{Gong:2014qga} 
See e.g. 
  J.~O.~Gong,
  %``Blue running of the primordial tensor spectrum,''
  JCAP {\bf 1407}, 022 (2014)
  [arXiv:1403.5163 [astro-ph.CO]].
  %%CITATION = ARXIV:1403.5163;%%


\bibitem{gsr}
%\cite{Stewart:2001cd}
%\bibitem{Stewart:2001cd} 
  E.~D.~Stewart,
  %``The Spectrum of density perturbations produced during inflation to leading order in a general slow roll approximation,''
  Phys.\ Rev.\ D {\bf 65}, 103508 (2002)
  [astro-ph/0110322]~;
  %%CITATION = ASTRO-PH/0110322;%%
%\cite{Choe:2004zg}
%\bibitem{Choe:2004zg} 
  J.~Choe, J.~O.~Gong and E.~D.~Stewart,
  %``Second order general slow-roll power spectrum,''
  JCAP {\bf 0407}, 012 (2004)
  [hep-ph/0405155].
  %%CITATION = HEP-PH/0405155;%%


\end{thebibliography}
\end{document}